\documentclass[fleqn,usenatbib]{mnras}

\usepackage{newtxtext,newtxmath}

\usepackage[T1]{fontenc}

\newcommand{\fermi}{{\it Fermi}}

\newcommand{\gm}{$\gamma$}

\DeclareRobustCommand{\VAN}[3]{#2}
\let\VANthebibliography\thebibliography
\def\thebibliography{\DeclareRobustCommand{\VAN}[3]{##3}\VANthebibliography}

\usepackage{graphicx}	
\usepackage{amsmath}	

\title[\fermi-BCU]{Optical spectroscopic characterization of \fermi~blazar candidates of uncertain type with TNG and DOT: First Results}

\author[Olmo-Garc\'{i}a et al.]{
Amanda Olmo-Garc\'{i}a,$^{1,2}$
Vaidehi S. Paliya,$^{3}$\thanks{E-mail: vaidehi.s.paliya@gmail.com (VSP)}
Nuria \'Alvarez Crespo$^{2,4}$\thanks{E-mail: nalvarezcrespo@ucm.es (NAC)}
Brajesh Kumar,$^{5}$
\newauthor
Alberto Dom{\'{\i}}nguez,$^{2,4}$
Armando Gil de Paz$^{1,2}$
and Patricia S\'anchez-Bl\'azquez$^{1,2}$
\\
$^{1}$Departamento de F\'{i}sica de la Tierra y Astrof\'{i}sica, Universidad Complutense de Madrid, E-28040 Madrid, Spain\\
$^{2}$Instituto de F\'isica de Part\'iculas y del Cosmos (IPARCOS), Universidad Complutense de Madrid, E-28040 Madrid, Spain\\
$^{3}$Inter-University Centre for Astronomy and Astrophysics (IUCAA), SPPU Campus, 411007, Pune, India\\
$^{4}$Departamento de Estructura de la Materia, F\'isica T\'ermica y Electr\'onica, Universidad Complutense de Madrid, E-28040 Madrid, Spain\\
$^{5}$Aryabhatta Research Institute of Observational Sciences (ARIES), Manora Peak, Nainital 263001, India
}

\date{Accepted XXX. Received YYY; in original form ZZZ}

\pubyear{2021}

\begin{document}
\label{firstpage}
\pagerange{\pageref{firstpage}--\pageref{lastpage}}
\maketitle

\begin{abstract}
The classification of \gm-ray-detected blazar candidates of uncertain type (BCU) is a relevant problem in extragalactic \gm-ray astronomy. Here we report the optical spectroscopic characterization, using two 3-4~m class telescopes, Telescopio Nazionale Galileo and Devasthal Optical Telescope, of 27 BCUs detected with the \fermi~Large Area Telescope. Since the identification of emission lines is easier in broad-line blazars, which usually exhibit low frequency peaked (synchrotron peak frequency $\leqslant10^{14}$ Hz) spectral energy distribution, we primarily target such BCUs. We found that 8 out of 27 sources exhibit broad emission lines in their optical spectra, 3 of them have redshifts $>$1 and the farthest one is at $z=2.55$. The optical spectra of 2 of the 19 remaining objects are dominated by the absorption spectra of the host galaxy, and there is a tentative detection of the Lyman-$\alpha$ absorption feature in one source. The spectra of the remaining 16 objects, on the other hand, are found to be featureless. 
\end{abstract}

\begin{keywords}
galaxies: active -- gamma-ray: galaxies  -- BL Lacertae objects: general
\end{keywords}



\section{Introduction}

The Large Area Telescope (LAT) onboard the {\it Fermi Gamma-ray Space Telescope} has detected thousands of extragalactic objects in high-energy \gm-rays \citep[][]{2020ApJS..247...33A}. The majority of them ($>$90\%) are blazars, an extreme class of active galactic nuclei (AGN) with relativistic jets aligned close to the line of sight to the observer \citep[][]{2020ApJ...892..105A}. Blazars are distinguished into two classes: BL Lac objects having emission lines in their optical spectra with rest-frame equivalent width EW$\leqslant$5 \AA, and Flat Spectrum Radio Quasars (FSRQs, EW$>$5 \AA), showing optical features as in normal quasar spectra \citep[][]{1991ApJ...374..431S}. The broadband spectral energy distribution (SED) of blazars shows a typical double-peaked structure with the lower energy peak explained by the synchrotron process and the high-energy emission originating via the inverse Compton mechanism. FSRQs generally exhibit low frequency peaked SEDs (synchrotron peak frequency $\nu_{\rm syn}^{\rm peak}\lesssim$10$^{14}$ Hz), whereas BL Lacs typically peak at higher frequencies \citep[$\nu_{\rm syn}^{\rm peak}>$10$^{15}$ Hz,][]{2020ApJ...892..105A}.

There are $\sim$30 per cent of \fermi-LAT detected sources that exhibit multi-frequency behavior similar to blazars (e.g., flat radio spectra), but often no optical spectra are available to identify their class (FSRQ/BL Lacs) or their signal-to-noise ratio is too low for a precise determination of their nature. Such objects are termed as blazar candidates of uncertain type \citep[BCUs, e.g.,][]{2011ApJ...743..171A}. There have been various methods developed to properly assess the physical properties of BCUs and classify them, e.g., based on their location in the WISE color-color diagram where \gm-ray emitting blazars occupy a distinct place \citep[][]{2012ApJ...750..138M} and machine learning algorithms \citep[cf.][]{2014ApJ...782...41D,2019ApJ...872..189K,2020MNRAS.493.1926K,2021JHEAp..29...40C}. However, none of these methods offer a conclusive way to clarify the nature of BCUs, unless there is an optical spectroscopic confirmation. Various optical spectroscopic followup campaigns have been organized to address this problem \citep[e.g.,][]{2014AJ....147..112P, 2015ANA...575A.124M, 2015AJ....149..163L, 2016AJ....151...95A, 2017MNRAS.467.2537K, 2017ApJ...851..135P, 2018AJ....156..212M, 2019ApJS..241....5D, 2019Ap&SS.364....5M, 2019Ap&SS.364...85P, 2019ApJ...871..162P, 2020ApJ...903L...8P,2021ApJS..254...26R}.

The classification of BCUs is crucial to (1) improve the completeness of \fermi-LAT catalogs, (2) build the luminosity function of blazars \citep[][]{2012ApJ...751..108A}, (3) select potential targets for the Cherenkov Telescope Array \citep[e.g.,][]{2013APh....43..215S,2017ICRC...35..632H,2021MNRAS.508.6128P}, (4) study emission mechanisms and physical properties \citep[e.g.,][]{2017ApJ...851...33P,2020ApJ...897..177P,2021ApJS..253...46P}, (5) obtain stringent limits on the extragalactic background light \citep[e.g.,][]{2018Sci...362.1031F,2019ApJ...874L...7D,2019A&A...629A...2F,2021MNRAS.507.5144S},  including constraints on the Hubble constant \citep[][]{2019ApJ...885..137D}, (6) search for counterparts of new flaring \gm-ray sources \citep[][]{2013A&A...551L...5B,2019RNAAS...3...92P}, (7) test new \gm-ray detection algorithms \citep[cf.][]{2015Ap&SS.360...65C}, (8) discover the new subclass of \gm-ray emitting sources, such as radio weak or extreme BL Lacs \citep[][]{2017ApJ...834..113M,2022MNRAS.512..137N}, and (9) search for electromagnetic counterparts of IceCube detected neutrinos \citep[e.g.,][]{2020ApJ...902...29P}.

Motivated by the above-mentioned science problems, we have started an optical spectroscopic follow-up campaign to secure optical spectra of reasonably bright \fermi-BCUs with medium 3-4 m class telescopes. In this paper, we present the results obtained for 27 blazars using  observations taken with 3.6 m Telescopio Nazionale Galileo (TNG) located at the Roque de Los Muchachos Observatory on the island of La Palma in the Canary Islands, Spain, and Devasthal Optical Telescope (DOT) located at the Devasthal Observatory, Nainital, India. It is a well-known fact that FSRQs exhibit broad emission lines which are easier to identify in relatively short exposure observations. Therefore, to identify potential FSRQs among the sample of BCUs, we prioritized sources with $\nu_{\rm syn}^{\rm peak}<$10$^{14}$ Hz. This is because broad emission line blazars or FSRQs are usually low synchrotron peaked (LSP) sources \citep[][]{2011ApJ...743..171A}, though some BL Lac sources could also exhibit LSP SEDs. This is shown in Figure~1. Additionally, we also observed  BL Lac 4FGL J1330.4+3157, associated with the radio source CGRaBS J1329+3154, which is reported to be a $z=3.79$ quasar based on its low signal-to-noise spectrum taken with the Sloan Digital Sky Survey \citep[][]{2020ApJS..250....8L}, to ascertain its spectroscopic redshift. We describe the sample selection in Section~\ref{sec:sample} and provide steps of data reduction in Section~\ref{sec:red}. The results are presented in Section~\ref{sec:results} followed by a brief description of individual BCUs in Section~\ref{sec:indiv}. We summarize our findings in Section~\ref{sec:summary}.

\begin{figure}
	\includegraphics[width=\linewidth]{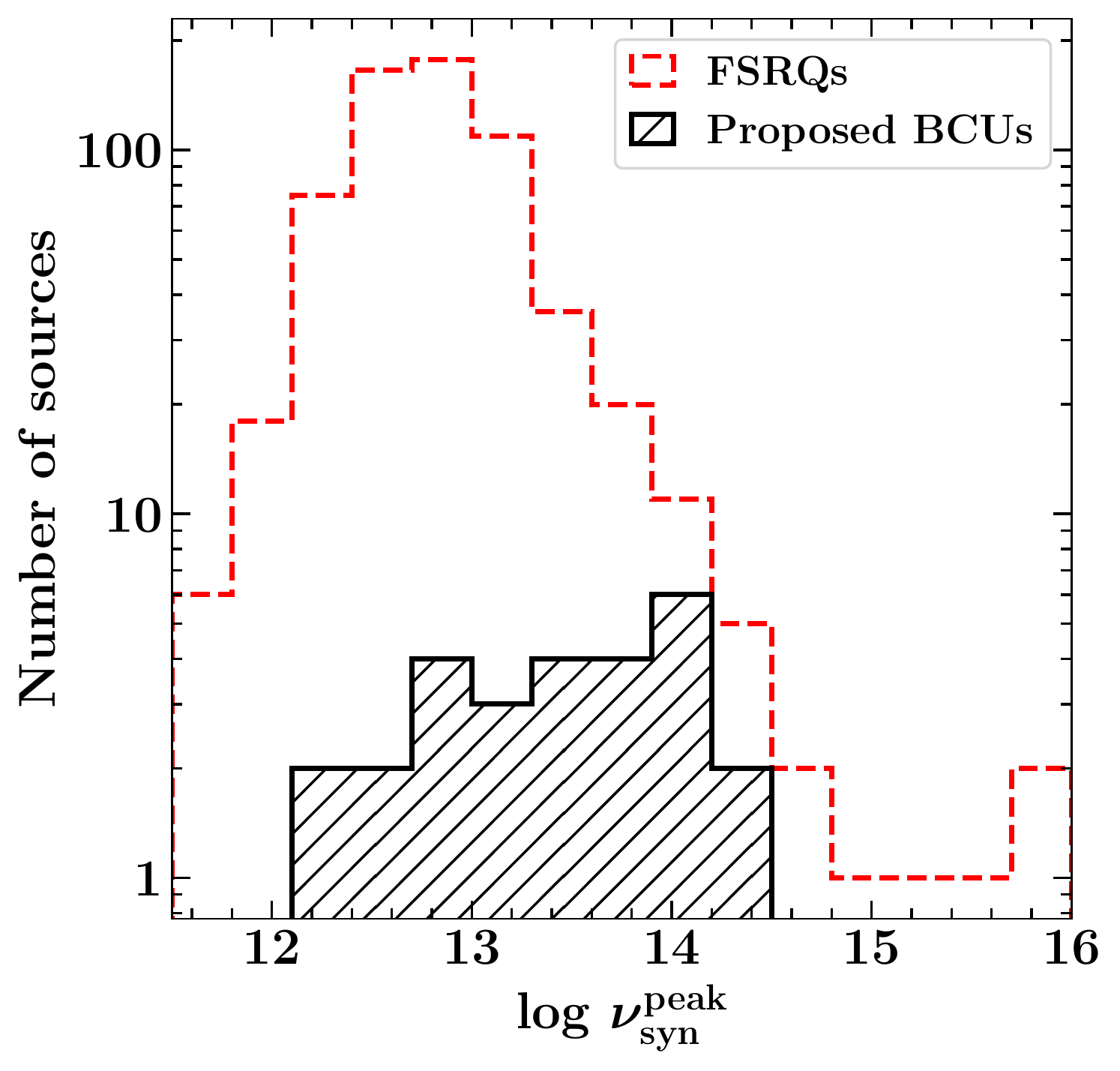}
    \caption{Distributions of the synchrotron peak frequencies.}
    \label{fig:nu_pk}
\end{figure}

\section{Sample Selection}\label{sec:sample}
The second data release of the fourth catalog of the \fermi~LAT detected AGN \citep[4LAC-DR2;][]{2020ApJ...892..105A} contains 1499 BCUs including 1292 blazars with no redshift information. All of them have low-frequency, usually radio, counterparts reported in 4LAC-DR2. The association of a \gm-ray emitter with a radio/optical/X-ray source is usually done following statistical methods as adopted in \fermi-LAT catalogs \citep[cf.][]{2020ApJ...892..105A}. For example, the likelihood ratio association technique \citep[][]{1996ApJ...461..396M,2011ApJ...743..171A} considers several surveys, e.g., NRAO VLA Sky Survey \citep[NVSS,][]{1998AJ....115.1693C} and the ROSAT All Sky Survey \citep[][]{1999A&A...349..389V}, and estimates the probability that a candidate is a `true' counterpart. The angular separation and brightness of radio sources lying within the considered uncertainty region of the \gm-ray source are taken into account while deriving the association probability. For a probability value $>$80\%, an object is considered as a high-confidence counterpart of the \gm-ray source \citep[see, further details in][]{2011ApJ...743..171A}.

We selected blazars which were visible from the TNG (28$^{\circ}$45'14.4"N, 17$^{\circ}$53'17.3" W) and the DOT (29$^{\circ}$21'40"N; 79$^{\circ}$41'04" E) and having $R$-band magnitude $<$ 20 mag. At TNG, we were granted 4.5, 1.5 and 4.5 nights during cycle 2020B (January 9-13, 2021), 2021A (July 2-3, 2021) and 2021B (December 6-10, 2021), respectively. Additionally, we were allocated half service night at TNG (May 22, 2021). Due to bad weather, 4 of the allocated 4.5 nights and 2.5 of the allocated 4.5 nights in cycle 2020B and 2021B, respectively, were lost. On DOT, we were awarded one night in the 2021-C1 cycle. Altogether, we were able to acquire optical spectra of 27 \fermi-BCUs. The general properties of these objects are provided in Table~\ref{tab:basic_info}.

\begin{table*}
\caption{The list of the $\gamma$-ray detected BCUs monitored in this work. Column information are as follows: (1) 4FGL source name; (2) other name; (3) right ascension (J2000); (4) declination (J2000); (5) apparent $R$-band magnitude; (6) Galactic extinction ($E(B-V)$) in mag; (7) date of observation (yyyy/mm/dd); (8) telescope used; (9) exposure time, in seconds; and (10) mean signal-to-noise ratio of the spectrum.\label{tab:basic_info}
}
\begin{center}
\begin{tabular}{llcccccccc}
\hline
4FGL Name  & Assoc. Name & RA (J2000) & Decl. (J2000) & $R$ mag. & $E(B-V)$~$^a$ & Date & Telescope & Exposure & S/N\\
(1) & (2) & (3)  & (4) & (5) &  (6)  & (7) & (8) & (9) & (10)\\ 
\hline															     			
J0024.4+4647   & B3 0021+464             & 00 24 21.54 & +46 44 06.2   & 19.07 & 0.07 & 2021/12/10 & TNG & 3600 & 18\\
J0028.9+3553   & GB6 J0028+3550          & 00 28 51.97 & +35 50 36.1   & 18.35 & 0.06 &  2021/01/13 & TNG   &  2400 & 11\\
J0130.6+1844   & MG1 J013030+1843        & 01 30 30.65 & +18 43 21.9   & 19.56 & 0.05 & 2021/12/09 & TNG & 5400 & 8\\
J0223.5+3912   & B3 0220+390             & 02 23 28.40 & +39 12 51.3   & 18.51 & 0.05 & 2021/12/10 & TNG & 2400 & 48\\
J0233.5+0654   & TXS 0230+067            & 02 33 29.99 & +06 55 26.4   & 18.76 & 0.11 & 2021/01/13  & TNG   &  2400 & 10\\
J0244.7+1316   & GB6 J0244+1320          & 02 44 45.69 & +13 20 07.2   & 18.91 & 0.09 & 2021/01/13  & TNG   &  2400 & 9\\
J0421.0$-$0752 & PKS 0418$-$079          & 04 20 53.94 & $-$07 52 19.9 & 18.49 & 0.09 & 2021/12/09 & TNG & 2400 & 18\\
J0430.3+1654   & MG1 J043022+1655        & 04 30 22.35 & +16 55 04.7   & 18.79 & 0.32 & 2021/01/13  & TNG   &  2400 & 11\\
J0526.3+2246   & NVSS J052622+224801     & 05 26 22.03 & +22 48 02.1   & 19.31 & 0.56 & 2021/12/09 & TNG & 3600 & 6\\
J0604.9+0000   & GB6 J0604+0000          & 06 04 58.42 & +00 00 43.2   & 16.04 & 0.50 & 2021/12/09 & TNG & 1200 & 43\\
J0638.2+6020   & GB6 J0638+6016          & 06 38 35.75 & +60 17 03.0   & 18.53 & 0.07 & 2021/12/09 & TNG & 2400 & 13\\
J0704.7+4508   & B3 0701+451             & 07 04 50.97 & +45 02 41.8   & 18.50 & 0.09 & 2021/01/13  & TNG   &  2400 & 16\\
J0746.5$-$0710 & PMN J0746$-$0709        & 07 46 27.49 & $-$07 09 49.6 & 18.87 & 0.10 & 2021/03/21  & DOT   &  5400 & 7\\
J0749.6+1324   & GB6 B0746+1329          & 07 49 35.95 & +13 21 56.0   & 19.45 & 0.03 & 2021/12/10 & TNG & 4800 & 2\\
J0804.5+0414   & TXS 0802+043            & 08 04 43.52 & +04 14 10.0   & 18.41 & 0.02 & 2021/12/10 & TNG & 2400 & 19\\
J0929.6+4621   & GB6 B0926+4634          & 09 29 22.76 & +46 20 46.5   & 18.08 & 0.02 & 2021/12/09-10 & TNG & 2400 & 9\\
J0959.6+4606   & 2MASX J09591976+4603515 & 09 59 19.81 & +46 03 51.9   & 18.06 & 0.01 & 2021/03/21  & DOT   &  2400 & 19\\
J1118.2$-$0415 & PMN J1118$-$0413        & 11 18 12.46 & $-$04 13 24.4 & 19.15 & 0.04 & 2021/03/21  & DOT   &  6000 & 21\\
J1129.2$-$0529 & NVSS J112914$-$052856   & 11 29 14.06 & $-$05 28 56.3 & 18.68 & 0.04 & 2021/05/22  & TNG   &  2400 & 21\\
J1330.4+3157   & CGRaBS J1329+3154       & 13 29 52.87 & +31 54 11.06  & 20.00 & 0.01 & 2022/04/19  & TNG   & 5400 & 7\\
J1419.4$-$0838 & NVSS J141922$-$083830   & 14 19 22.56 & $-$08 38 32.1 & 18.13 & 0.04 & 2021/03/21  & DOT   &  3600  & 12\\
J1549.3+6310   & WN B1549+6319           & 15 49 57.32 & +63 10 07.3   & 17.75 & 0.02 & 2021/05/22   & TNG   &  1200 & 9\\
J1756.9+1531   & 87GB 175437.6+153548    & 17 56 53.10 & +15 35 20.8   & 19.36 & 0.09 &  2021/07/03 & TNG   &  3600 & 11\\
J1822.0+1600   & OU 134                  & 18 22 09.97 & +16 00 14.8   & 18.59 & 0.17 & 2021/07/02   & TNG   &  3600 & 39\\
J1836.5+1948   & NVSS J183632+195047     & 18 36 32.11 & +19 50 46.3   & 19.33 & 0.22 & 2021/07/02  & TNG   &  3600 & 22\\
J1845.0+1613   & 87GB 184225.9+161105    & 18 44 42.61 & +16 14 11.1   & 17.93 & 0.40 & 2021/07/02  & TNG   &  2400 & 21\\
J1949.0+1314   & 87GB 194635.4+130713    & 19 48 55.22 & +13 14 39.9   & 19.00 & 0.24 &  2021/07/03 & TNG   &  3600 & 29\\
\hline
\end{tabular}
\begin{tabular}{c}
\noalign{\smallskip} 
\textsc{Note} --- $^a$ Average reddening value by \citet{2011ApJ...737..103S}, obtained from \url{https://irsa.ipac.caltech.edu/applications/DUST/}. \\
\end{tabular}
\end{center}
\end{table*}

\section{Observations and Data Reduction}\label{sec:red}
\subsection{Telescopio Nazionale Galileo (TNG)}\label{sec:tng}
We used the spectrograph DOLORES (Device Optimized for the LOw RESolution) to take optical spectra of \fermi-BCUs with TNG. 
We used the  LR-B grism and the 1 arcsec slit, which gives a spectral resolution $R=585$, and a reciprocal dispersion of 2.52 \AA~pixel$^{-1}$ in the wavelength range 3000 to 8430 \AA. The slit was oriented at the parallactic angle. The total integration time was divided into exposures of 20 minutes as a compromise between signal-to-noise and contamination from cosmic rays. The total integration time for each object is listed in Table~\ref{tab:basic_info}.

\subsection{Devasthal Optical Telescope (DOT)}\label{sec:dot}
We used the low resolution slit-spectrograph ARIES Devasthal Faint Object Spectrograph \& Camera (ADFOSC) mounted at the 3.6 m DOT.
The spectral dispersions provided by the different grisms are in the range of 1 to 7 \AA~pixel$^{-1}$ and an 8$\arcmin$ long slit with different available widths (0.4$\arcsec$--3.2$\arcsec$) can be used. The spectra cover the wavelength range 3500-10500 \AA with a spectral resolution $R=515$. Further details about the instrument can be found in \citet[][]{2012SPIE.8446E..14O}. We observed 5 \fermi-BCUs with ADFOSC on the night of March 21, 2021 (Table~\ref{tab:basic_info}).

\subsection{Data Reduction}
The reduction of the spectra was done following the standard procedure with  IRAF\footnote{\url{http://iraf.noao.edu/} IRAF is distributed by the National Optical Astronomy Observatory, which is operated by the Association of Universities for Research in Astronomy (AURA) under cooperative agreement with the National Science Foundation.} tasks, through the PyRAF software\footnote{PyRAF is a product of the Space Telescope Science Institute, which is operated by AURA for NASA \url{http://www.stsci.edu/institute/software hardware/pyraf/}}. The initial steps are: bias subtraction, flat correction, cosmic ray removal \citep{2001PASP..113.1420V} and wavelength calibration. Flats were not available in the data from telescope DOT. Since this step corrects pixel to pixel differences in sensitivity, the reduced spectra will be slightly noisier, although subsequent steps in the data reduction will  mitigate it. In the wavelength calibration, we used a combination of arc lamps: Ne+Hg and He for TNG and Hg+Ar for DOT.

The next step is the sky subtraction. We took samples in each side of the object and fitted a polynomial of degree 1 or 2, when needed, and subtracted it from the spectra. After the sky subtraction, we used  the fitted sky to check the wavelength calibration. We measured the wavelength of emission lines in the fitted sky background. In the objects where we found a constant offset between the measured and  the reference value, we subtracted  it from the wavelength solution. 

Afterwards, we did the combination  and  extraction of the spectra and the flux calibration. Some objects improved their signal to noise ratio when reversing the order, extracting first and combining in the end. For the flux calibration, we observed spectrophotometric standard stars in each night of observation. We also corrected the atmospheric extinction at the observatory and the Galactic extinction with the values R=3.1 and $E(B-V)$ described in Table~\ref{tab:basic_info} \citep{2011ApJ...737..103S}. 

\begin{figure*}
    \hbox{\hspace{2.cm}
	\includegraphics[scale=0.5]{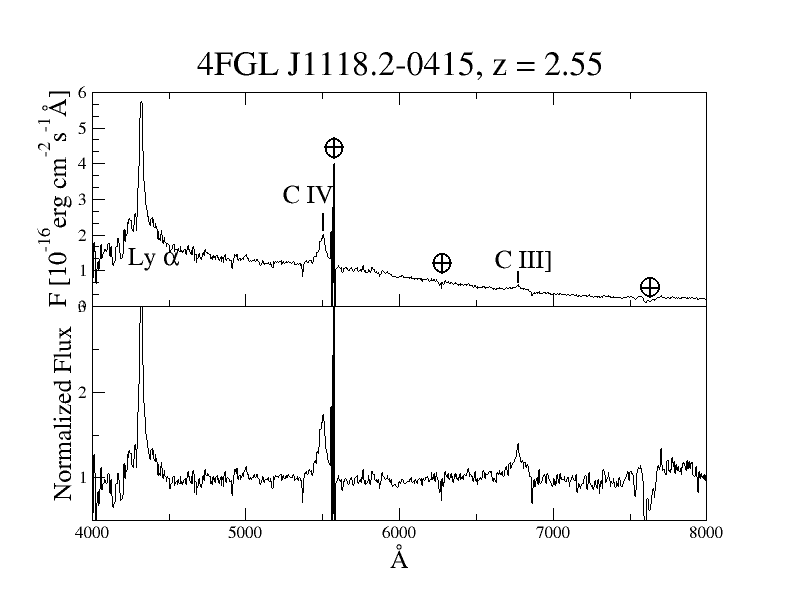}
	}
    \caption{Optical spectrum of the BCU 4FGL J1118.2$-$0415 taken with the DOT. Top panel shows the observed spectrum whereas the normalized spectrum is shown in the bottom panel. Absorption/emission lines are highlighted. $\oplus$ symbol indicates the position of sky emission and telluric absorption features.}
    \label{fig:A1}
\end{figure*}

\begin{table*}
\caption{Table with the emission/absorption lines identified in each spectrum. Column information are as follows: (1) 4FGL source name; (2) redshift derived from the identified lines; and (3) name of the identified lines.\label{tab:lines}
}
\begin{center}
\begin{tabular}{lccc}
\hline
4FGL Name   &   Redshift & Classification & Lines identified \\
(1) & (2) & (3)  & (4)\\ 
\hline		
J0024.4+4647    & 1.91  & FSRQ    & Ly-$\alpha$, Si\,{\sc iv}, C\,{\sc iv}, C\,{\sc iii} \\
J0028.9+3553    &       & BL Lac  &  \\
J0130.6+1844    &       & BL Lac  &  \\
J0223.5+3912    &       & BL Lac  &  \\
J0233.5+0654    &       & BL Lac  &  \\
J0244.7+1316    & 0.98  & FSRQ    & Mg\,{\sc iv}, [Ar\,{\sc iv}] \\
J0421.0$-$0752  & 0.28  & BL Lac  & Ca\,{\sc ii} H\,\&\,K, G band, Mg\,{\sc i} \\ 
J0430.3+1654    &       & BL Lac  &  \\
J0526.3+2246    &       & BL Lac  &  \\
J0604.9+0000    & 0.30  & BL Lac  & Ca\,{\sc ii} H\,\&\,K, H$\delta$, G-band, Mg\,{\sc i}\\
J0638.2+6020    &       & BL Lac  &  \\
J0704.7+4508    &       & BL Lac  &  \\
J0746.5$-$0710  & 0.89  & FSRQ    & Mg\,{\sc ii} \\
J0749.6+1324    & 1.05  & FSRQ    & C\,{\sc iii}, Mg\,{\sc ii}, [Ne\,{\sc iii}]\\
J0804.5+0414    &       & BL Lac  &  \\
J0929.6+4621    &       & BL Lac  &  \\
J0959.6+4606    & 0.14  & FSRQ    &  G-band, [O\,{\sc iii}], Mg\,{\sc i}, Na-D\\
J1118.2$-$0415  & 2.55  & FSRQ    & Ly-$\alpha$ C\,{\sc iv}, C\,{\sc iii}\\
J1129.2$-$0529  &       & BL Lac  &  \\
J1330.4+3157    & 2.01  & BL Lac  & Ly-$\alpha$\\
J1419.4$-$0838  & 0.90  & FSRQ    & Mg\,{\sc ii}  \\ 
J1549.3+6310    &       & BL Lac  &  \\
J1756.9+1531    &       & BL Lac  &  \\
J1822.0+1600    & 0.98  & FSRQ    & C\,{\sc iii}, Mg\,{\sc ii}, [O\,{\sc ii}], H$\epsilon$ \\
J1836.5+1948    &       & BL Lac  &  \\
J1845.0+1613    &       & BL Lac  &  \\
J1949.0+1314    &       & BL Lac  &  \\
\hline
\end{tabular}
\end{center}
\end{table*}
\section{Results}\label{sec:results}
The reduced spectrum of one of the targets is shown in Figure~\ref{fig:A1}, whereas, the remaining are presented in different figures in Appendix~\ref{sec:appendix_fig}. We have also plotted the normalized spectrum, which is calculated by fitting the continuum with a cubic spline function and dividing the spectrum by the fitted continuum. The spectra were scanned to identify emission/absorption lines to estimate the source redshift. For a consistency check, we also inspected the individual exposures and verified whether lines are visible in all of them. Altogether, the redshifts were measured for 11 BCUs out of 27 sources. Following the conventional emission line EW-based blazar classification scheme, 8 objects are identified as FSRQs (EW$>$5\AA). The spectra of 2 BL Lac sources were found to be dominated by the absorption features arising from the host galaxy, e.g., Ca~{\sc ii} H\&K doublet, allowing us to determine their redshifts. We tentatively identified the Ly-$\alpha$ absorption feature in the low-S/N optical spectrum of the candidate high-redshift BL Lac source 4FGL J1330.4+3157 and derived the source redshift to be $z=2.01$. On the other hand, the featureless optical spectra of the remaining 16 BCUs imply that they can be classified as BL Lac objects. We discuss our findings about individual sources in the next section.

\section{Notes on Individual Objects}\label{sec:indiv}
\noindent{\it 4FGL J0024.4+4647}: The spectrum is characterized by a blue continuum and four emission lines, Ly-$\alpha$ (EW$_{\rm obs}$=148 \AA), Si\,{\sc iv} (EW$_{\rm obs}$=18.61 \AA), C\,{\sc iv} (EW$_{\rm obs}$=62.05 \AA), and C\,{\sc iii} (EW$_{\rm obs}$=21.2 \AA), were identified. The derived redshift is 1.91 and the source can be classified as a FSRQ.

\noindent{\it 4FGL J0028.9+3553, 4FGL J0130.6+1844, 4FGL J0223.5+3912, 4FGL J0233.5+0654}: The spectra of these BCUs are characterized by a blue continuum. No emission or absorption lines have been identified, implying them to be classified as BL Lac objects.

\noindent{\it 4FGL J0244.7+1316}: The spectrum is characterized by a blue continuum. The emission lines Mg\,{\sc ii} (EW$_{\rm obs}$=34.23 \AA) and [Ar\,{\sc iv}] (EW$_{\rm obs}$=5.8 \AA) are identified. The derived redshift is 0.98 and the source can be classified as a FSRQ.

\noindent{\it 4FGL J0421.0$-$0752}: The spectrum is characterized by a red continuum and faint absorption features typically seen in the optical spectrum of elliptical galaxies, such as Ca\,{\sc ii} K (EW$_{\rm obs}$=4.46 \AA), Ca\,{\sc ii} H (EW$_{\rm obs}$=2.88 \AA), G-band (EW$_{\rm obs}$=5.38 \AA), and Mg\,{\sc i} (EW$_{\rm obs}$=6.02 \AA), are observed. It is a BZG, a BL Lac with a host galaxy dominated spectrum. 

\noindent{\it 4FGL J0430.3+1654, 4FGL J0526.3+2246}: No emission/absorption lines are identified in the spectra of these BCUs. They can be categorized as BL Lac objects.

\noindent{\it 4FGL J0604.9+0000}: The spectrum shows strong absorption lines Ca\,{\sc ii} K (EW$_{\rm obs}$=2.24 \AA), Ca\,{\sc ii} H (EW$_{\rm obs}$=0.83 \AA), H$\delta$ (EW$_{\rm obs}$=2.74 \AA), G-band (EW$_{\rm obs}$=101.47 \AA), and Mg\,{\sc i} (EW$_{\rm obs}$=105.80 \AA), characteristic of an elliptical galaxy. The redshift measured is $z=$0.30. It is possibly a BZG, a BL Lac with host galaxy dominated spectrum.

\noindent{\it 4FGL J0638.2+6020, 4FGL J0704.7+4508}:  The spectrum is characterized by red and blue continua, respectively, and no emission or absorption lines have been identified, implying these BCUs to be classified as BL Lac objects. 

\noindent{\it 4FGL J0746.5$-$0710}: The spectrum is characterized by a red continuum and faint emission lines. The broad emission line Mg\,{\sc ii} (EW$_{\rm obs}$=16.92 \AA) is identified, leading to a source redshift of 0.90 and enabling the source to be classified as a FSRQ.

\noindent{\it 4FGL J0749.6+1324}: The spectrum is characterized  by several emission lines: C\,{\sc iii} (EW$_{\rm obs}$=243.5 \AA), Mg\,{\sc ii} (EW$_{\rm obs}$=289.1 \AA), and [Ne\,{\sc iii}] (EW$_{\rm obs}$=48.34 \AA). The derived source redshift is 1.05 and the object can be classified as a FSRQ.

\noindent{\it 4FGL J0804.5+0414, 4FGL J0929.6+4621}: The spectra are characterized by featureless continua. Accordingly, both sources can be classified as BL Lac objects. 

\noindent{\it 4FGL J0959.6+4606}: The spectrum is characterized by several faint absorption and narrow emission lines: G-band (EW$_{\rm obs}$= 8.9 \AA), [O\,{\sc iii}] (EW$_{\rm obs}$= 5.88 \AA), Mg\,{\sc i} (EW$_{\rm obs}$= 6.09 \AA), and Na-D (EW$_{\rm obs}$=4.93 \AA). Interestingly, the equivalent width of the [O\,{\sc iii}]5007 \AA~line is larger than 5 \AA, indicating this source to be likely a FSRQ at redshift $z=$ 0.14.

\noindent{\it 4FGL J1118.2-0415}: The spectrum is characterized by a blue continuum and strong emission lines: Ly-$\alpha$ (EW$_{\rm obs}$=95.46 \AA), C\,{\sc iv} (EW$_{\rm obs}$=26.54 \AA), and C\,{\sc iii} (EW$_{\rm obs}$=8.11 \AA). The object can be classified as a FSRQ at the redshift of 2.55. This is the most distant blazar observed in our sample.

\noindent{\it 4FGL J1129.2-0529}: The spectrum is featureless and the BCU can be classified as a BL Lac object.

\noindent{\it 4FGL J1330.4+3157}: An absorption trough at $\sim$3700 \AA\ is seen in the optical spectrum of this candidate high-redshift BL Lac object, which we interpret as Lyman-$\alpha$ absorption. Accordingly, the derived source redshift is $z=2.01$ thereby making it one of the very few $z>2$ \gm-ray detected BL Lac objects. A faint emission bump-like feature was also seen at $\sim$4400 \AA\ which could be due to C\,{\sc iv} emission supporting the source redshift of $z=2.01$. However, a strong claim cannot be made due to the faintness of the target. Observations from larger telescopes will be needed to better constrain the redshift.

\noindent{\it 4FGL J1419.4-0838}:  The spectrum is characterized  by a blue continuum and a strong Mg\,{\sc ii} (EW$_{\rm obs}$=40.72 \AA) emission line. The object can be classified as a FSRQ at the redshift of 0.90.


\noindent{\it 4FGL J1549.3+6310, 4FGL J1756.9+1531}: The spectra are characterized by featureless continua. Accordingly, the sources can be classified as BL Lac objects. 

\noindent{\it 4FGL J1822.0+1600}:  The spectrum is characterized by a blue continuum and strong emission lines: C\,{\sc iii} (EW$_{\rm obs}$=19.48 \AA), Mg\,{\sc ii} (EW$_{\rm obs}$=27.8 \AA), [O\,{\sc ii}] (EW$_{\rm obs}$=5.93 \AA), and H$\epsilon$ (EW$_{\rm obs}$=3.06 \AA). The object can be classified as a FSRQ at the redshift of 0.98.

\noindent{\it 4FGL J1836.5+1948, 4FGL J1845.0+1613, 4FGL J1949.0+1314}: The spectra of these BCUs are characterized by featureless continua. Accordingly, they can be classified as BL Lac objects.

\section{Conclusions}\label{sec:summary}
In this article, we have presented our findings based on the observations of 27 \gm-ray detected blazar candidates with the primary objective of measuring their redshift. We focused primarily on the LSP BCUs since almost all known broad emission line blazars are LSP sources and identifying emission lines in this class of objects is relatively easier and less telescope time consuming compared to BL Lac objects. Out of 27 sources, we found broad emission lines in 8 BCUs and narrow emission/absorption features in 2 objects. In our sample, there are 4 sources with redshifts $>$1. We admit that using only synchrotron peak frequency information may not be the only criterion to select potential broad-line blazars among the sample of \fermi-BCUs. Therefore, during upcoming observing runs within this long-term project, we will also consider other observational features, such as the location in the Wide-field Infrared Survey Explorer color-color diagram \citep[cf.][]{2012ApJ...750..138M} and Compton dominance information \citep[][]{2021ApJS..253...46P}. These extra conditions would enable us to choose targets having a larger probability of exhibiting broad emission lines.

\section*{Acknowledgements}

We are grateful to the journal referee for constructive criticism. Part of the data presented herein was obtained at the Devasthal Optical Telescope (DOT), which is operated by Aryabhatta Research Institute of Observational Sciences (ARIES). We thank the staff of ARIES who made these observations possible. Based on observations made with the Italian Telescopio Nazionale Galileo (TNG) operated by the Fundación Galileo Galilei (FGG) of the Istituto Nazionale di Astrofisica (INAF) at the Observatorio del Roque de los Muchachos (La Palma, Canary Islands, Spain). We thank the TNG telescope staff for the observations done in service mode. A.O.G. acknowledges financial support from the Spanish Ministry of Science, Innovation and Universities (MCIUN) under grant numbers RTI2018-096188-B-I00, and from the Comunidad de Madrid Tec2Space project S2018/NMT-4291. A.D. acknowledges the support of the Ram\'on y Cajal program from the Spanish MINECO. P.S-B acknowledges financial support from the Spanish Ministry of Science, Innovation and Universities (MCIUN) under grant number PID2019-107427GB-C31.

\section*{Data Availability}
The spectra from this article can be found at \url{https://www.ucm.es/blazars/BCUs}.


\bibliographystyle{mnras}
\bibliography{example} 


\appendix

\section{Optical Spectra} \label{sec:appendix_fig}
Here we show the optical spectra of all \fermi-BCUs observed with TNG and DOT.

\begin{figure*}
	\includegraphics[scale=0.25]{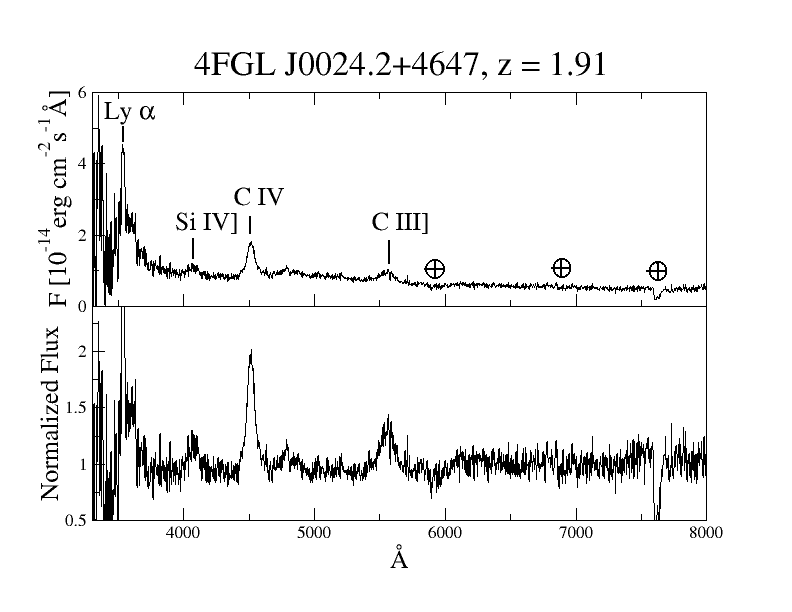}
	\includegraphics[scale=0.25]{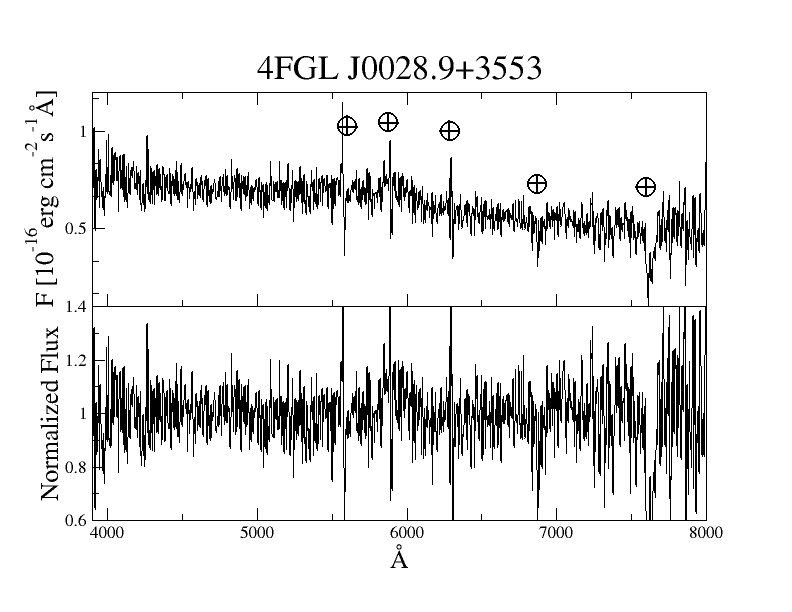}
	\includegraphics[scale=0.25]{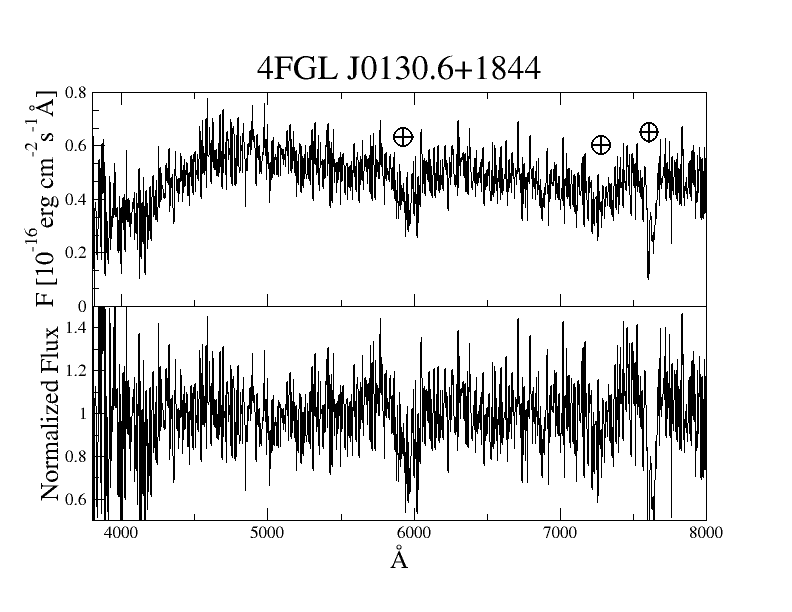}
	\includegraphics[scale=0.25]{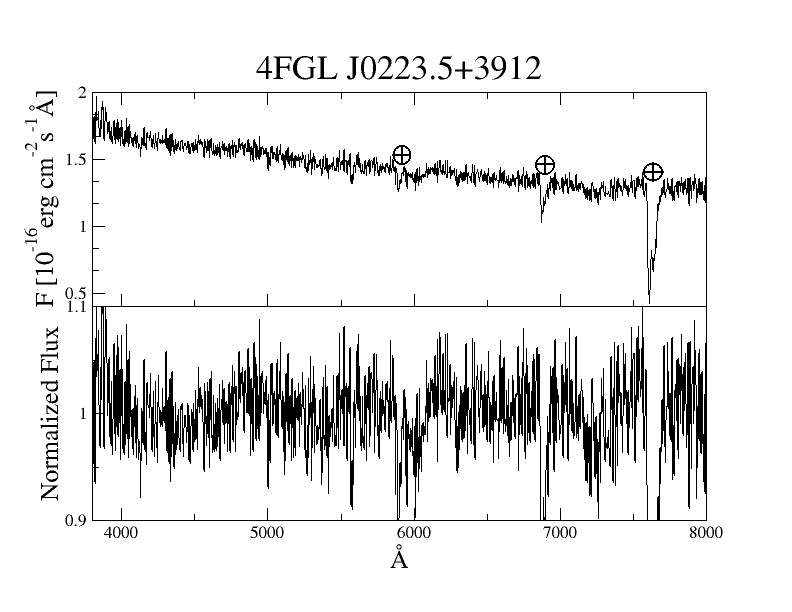}
	\includegraphics[scale=0.25]{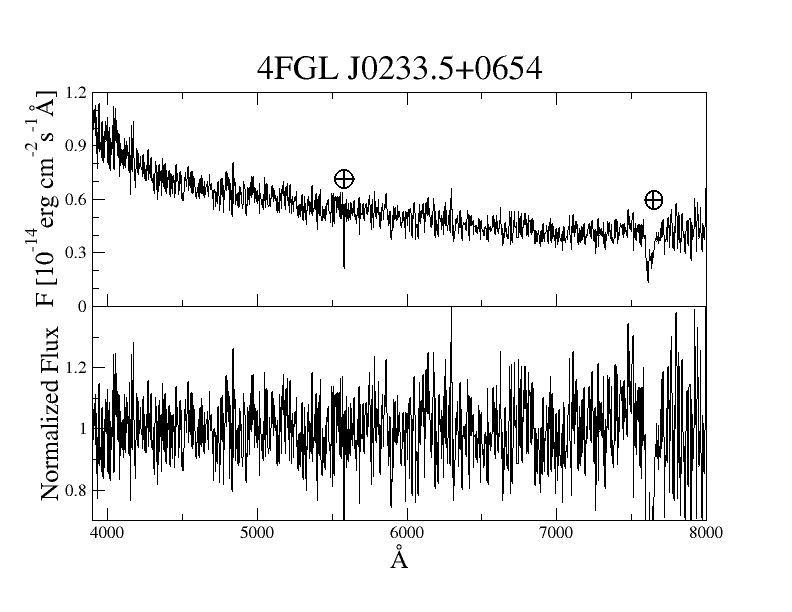}
	\includegraphics[scale=0.25]{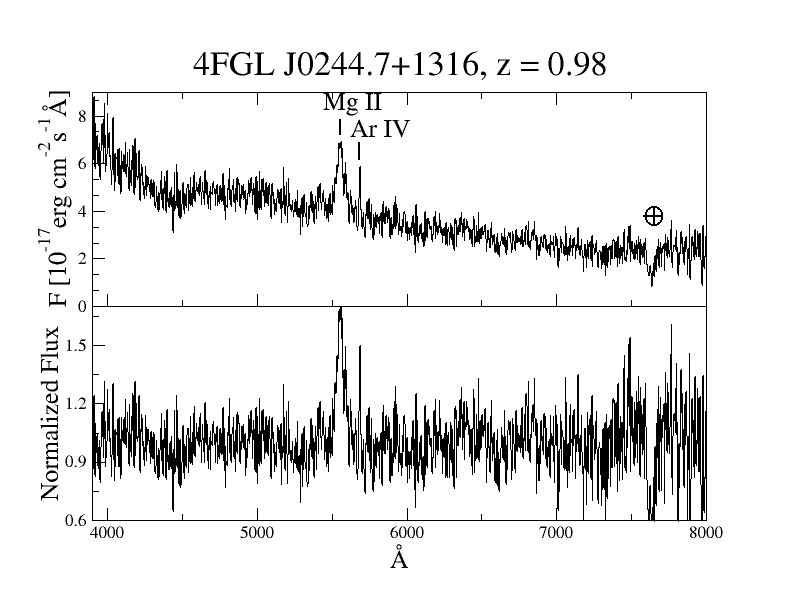}
	\includegraphics[scale=0.25]{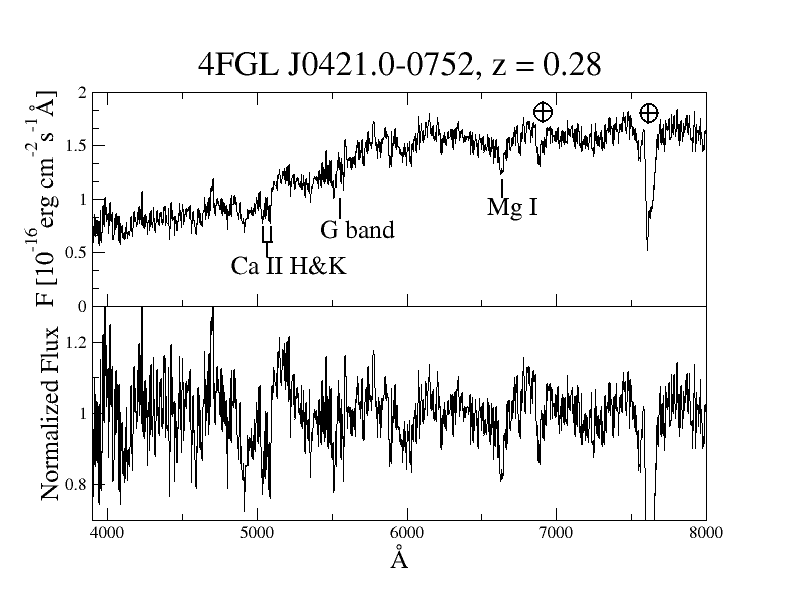}
	\includegraphics[scale=0.25]{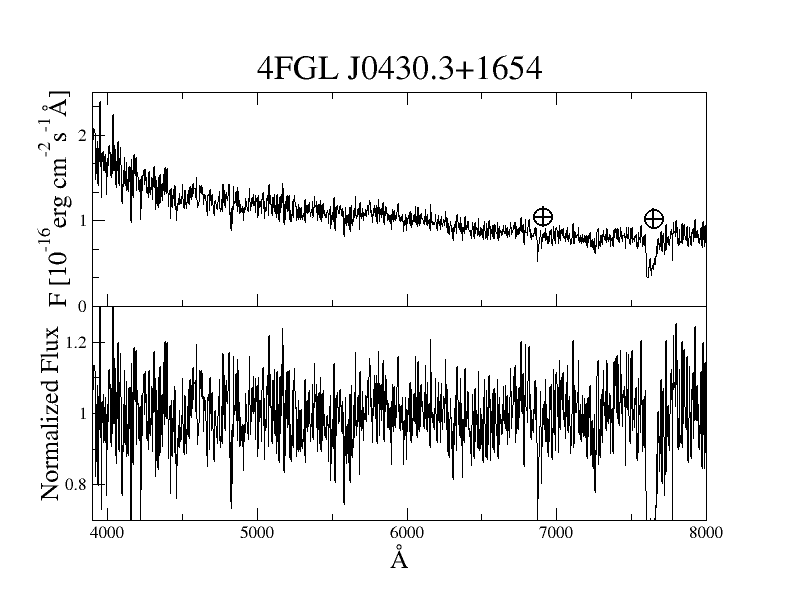}
    \caption{Same as Figure~\ref{fig:A1}}
    \label{fig:A2}
\end{figure*}
\begin{figure*}
	\includegraphics[scale=0.25]{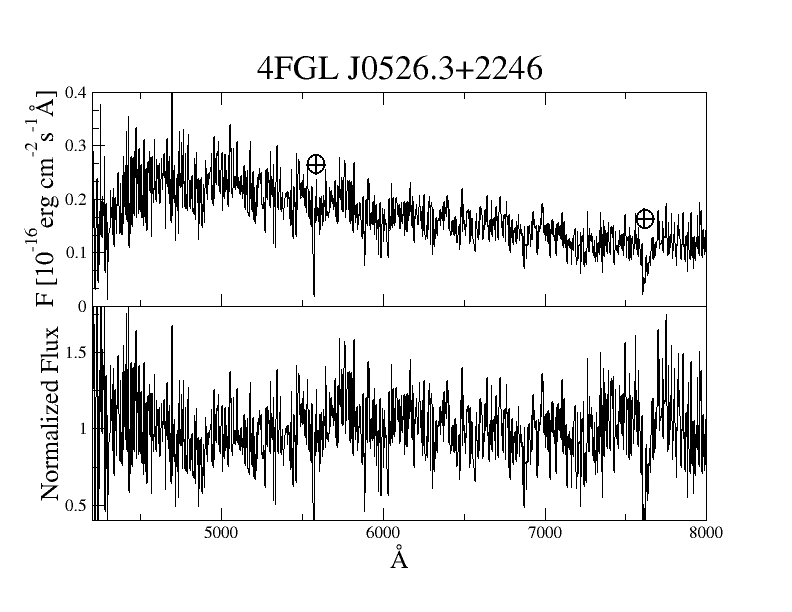}
	\includegraphics[scale=0.25]{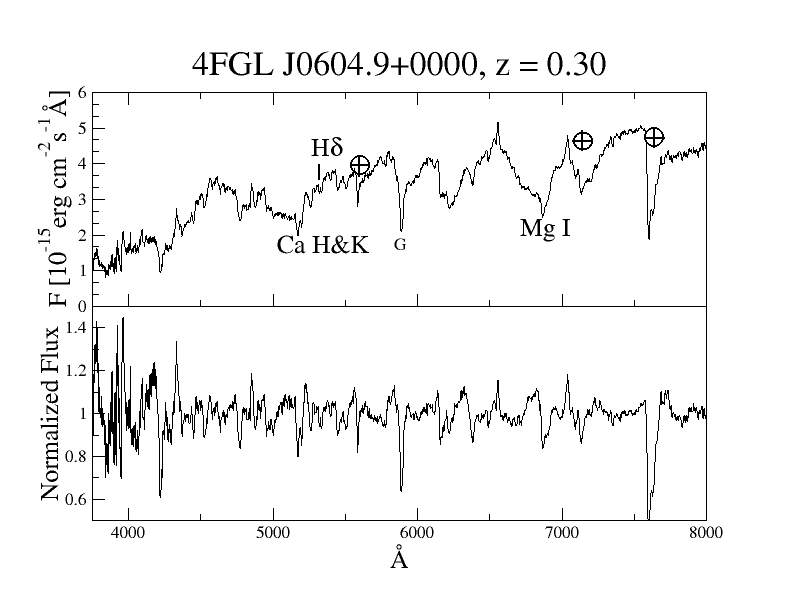}
	\includegraphics[scale=0.25]{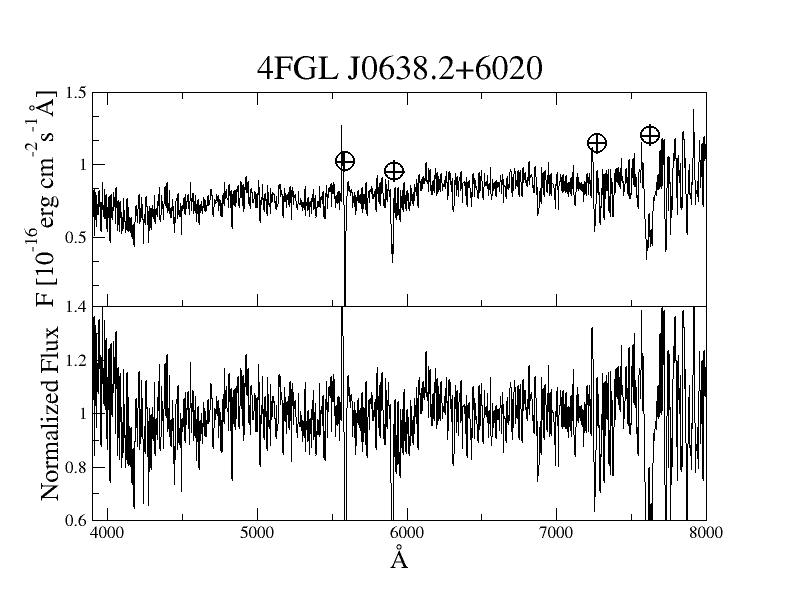}
	\includegraphics[scale=0.25]{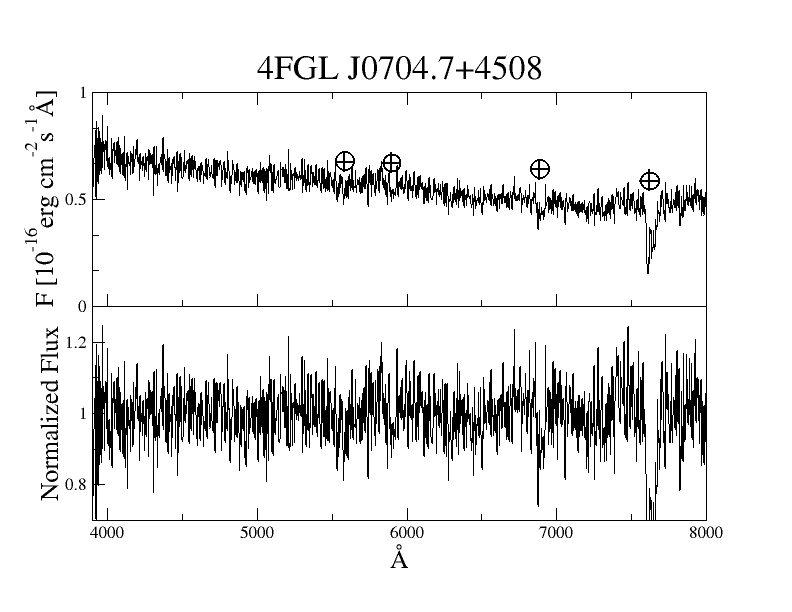}
	\includegraphics[scale=0.25]{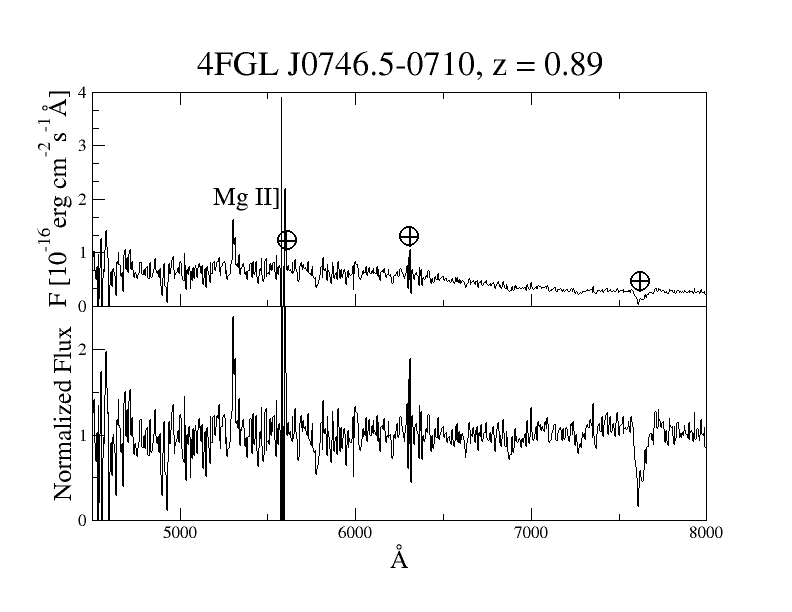}
	\includegraphics[scale=0.25]{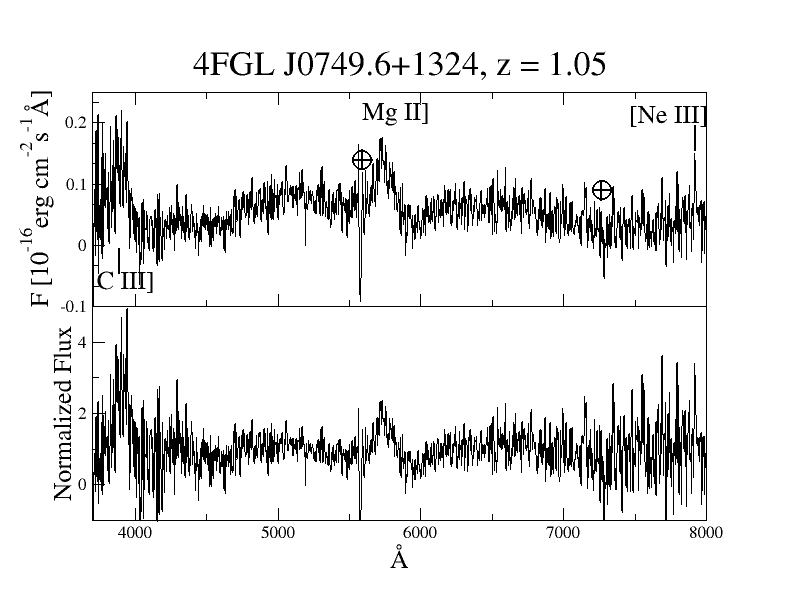}
	\includegraphics[scale=0.25]{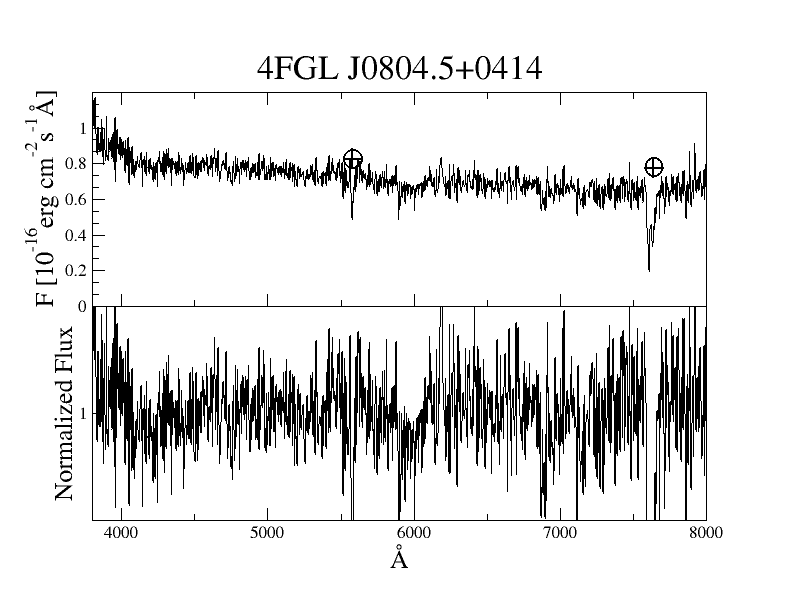}
	\includegraphics[scale=0.25]{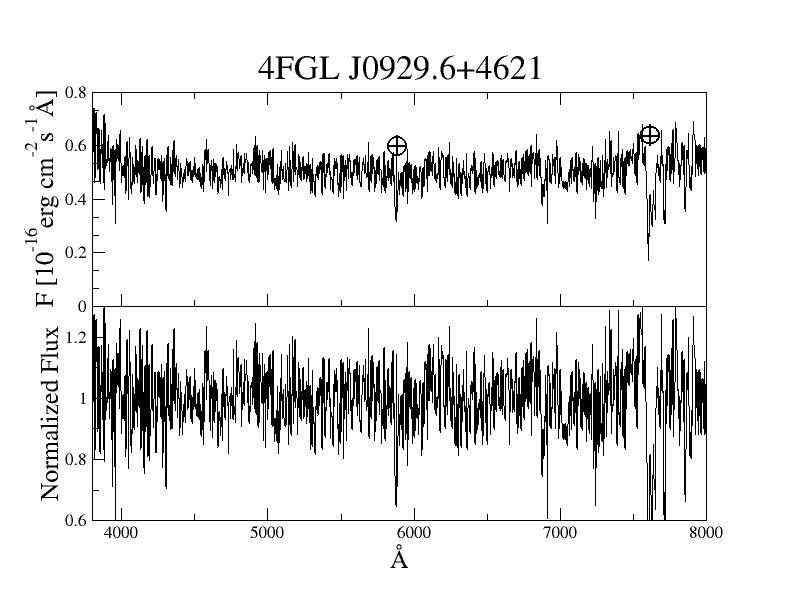}
	\caption{Same as Figure~\ref{fig:A1}.}
\end{figure*}
\begin{figure*}
	\includegraphics[scale=0.25]{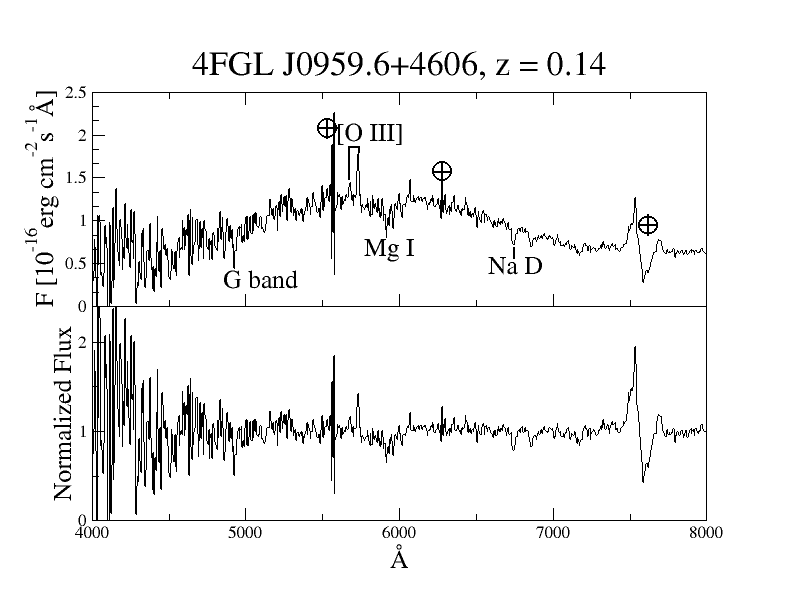}
	\includegraphics[scale=0.25]{J1118.png}
	\includegraphics[scale=0.25]{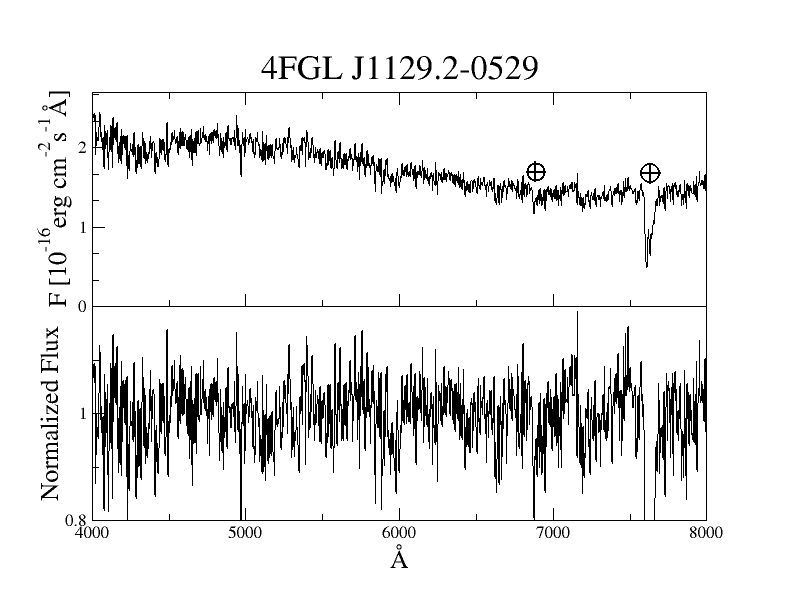}
	\includegraphics[scale=0.25]{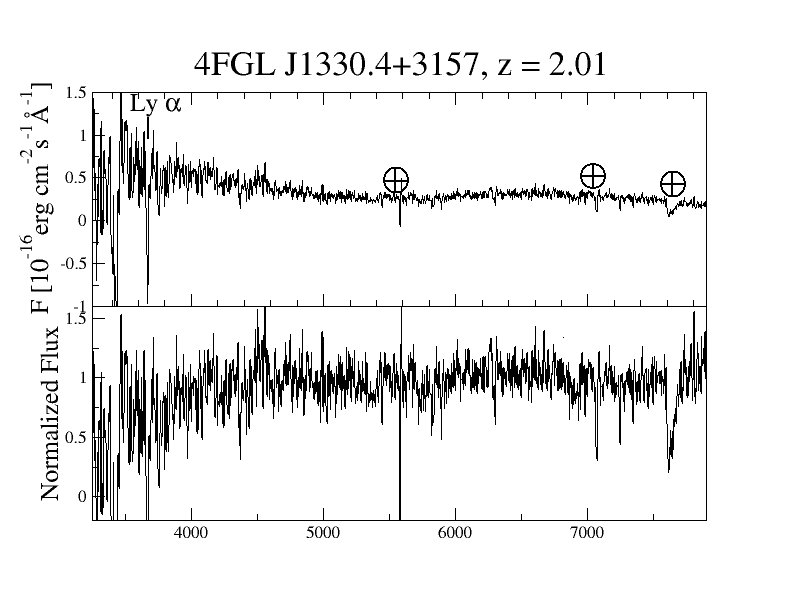}
	\includegraphics[scale=0.25]{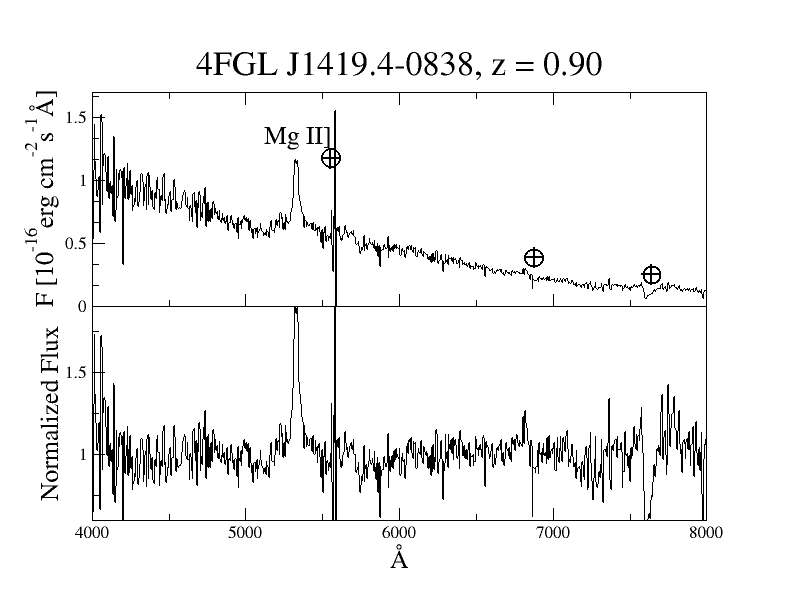}
	\includegraphics[scale=0.25]{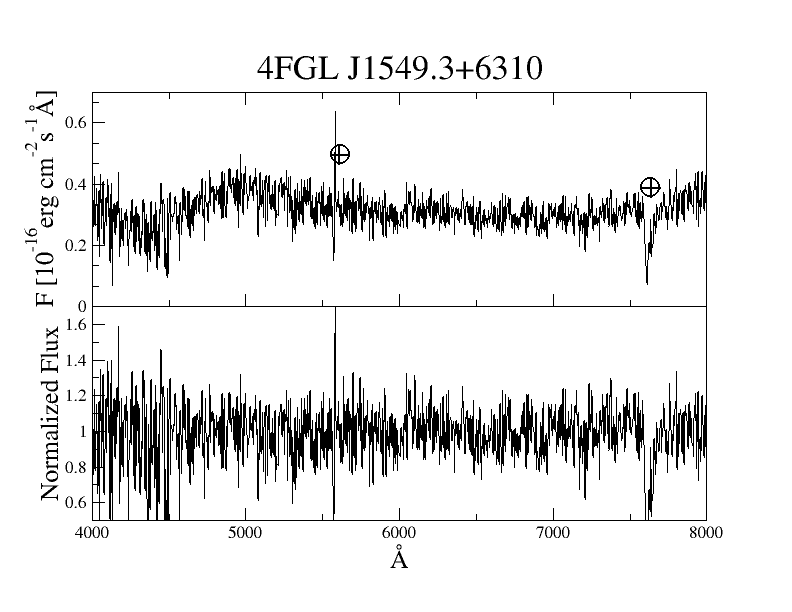}
        \includegraphics[scale=0.25]{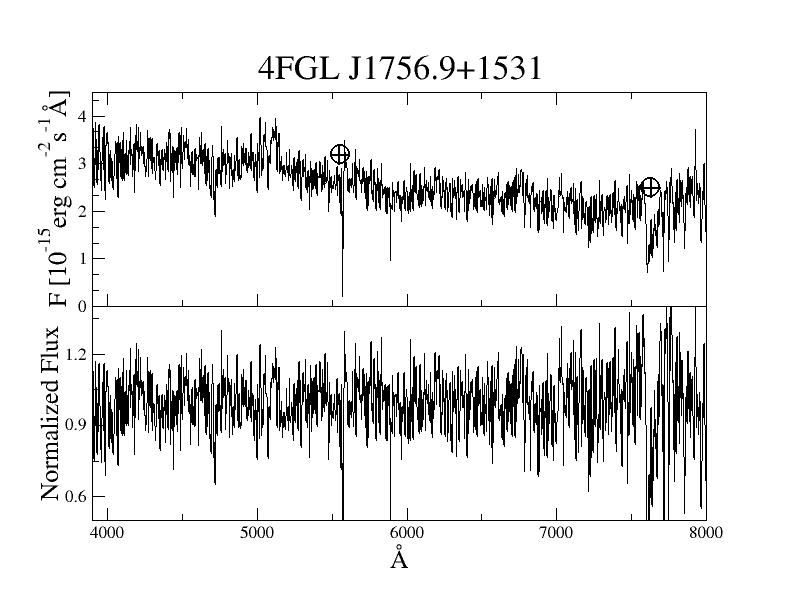}
	\includegraphics[scale=0.25]{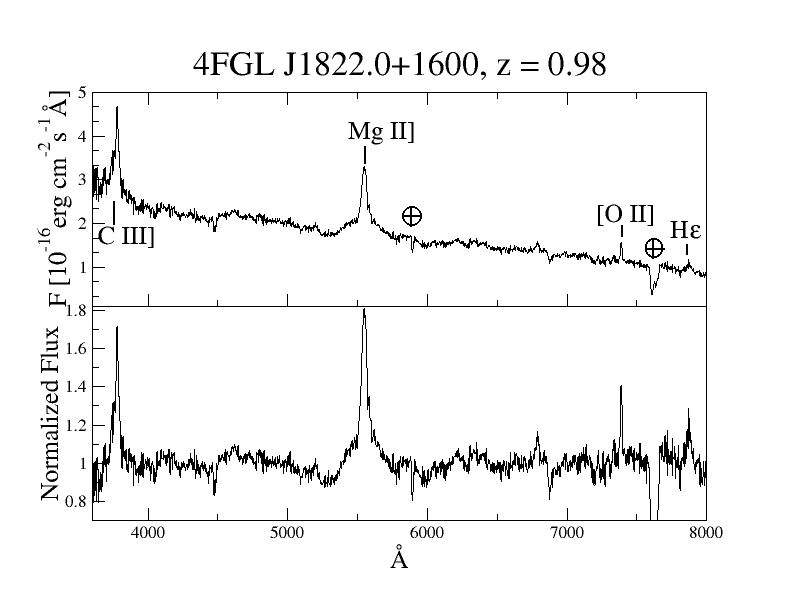}
	
	\caption{Same as Figure~\ref{fig:A1}.}
\end{figure*}
\begin{figure*}
	\includegraphics[scale=0.25]{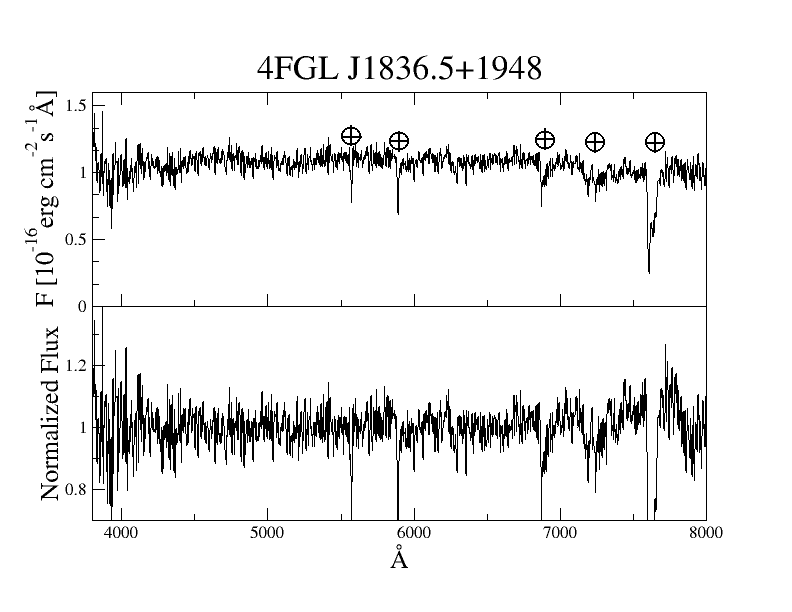}
	\includegraphics[scale=0.25]{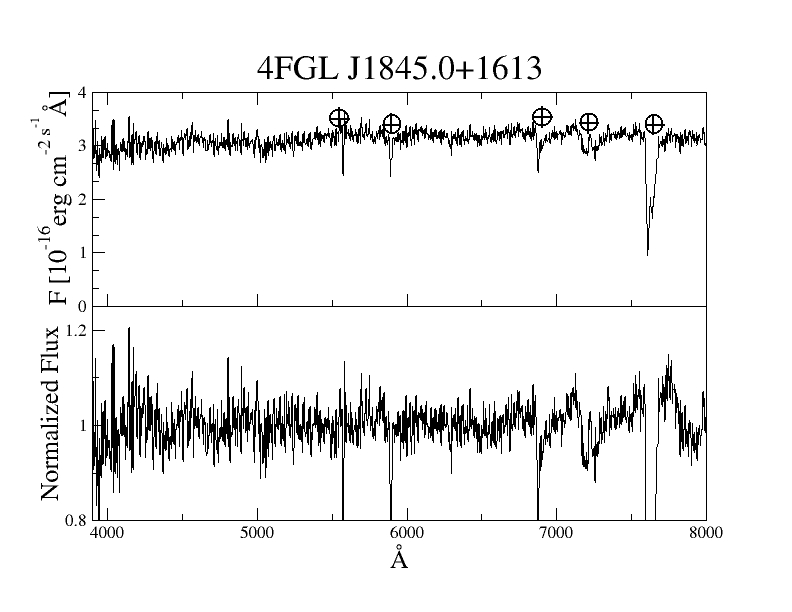}
	\includegraphics[scale=0.25]{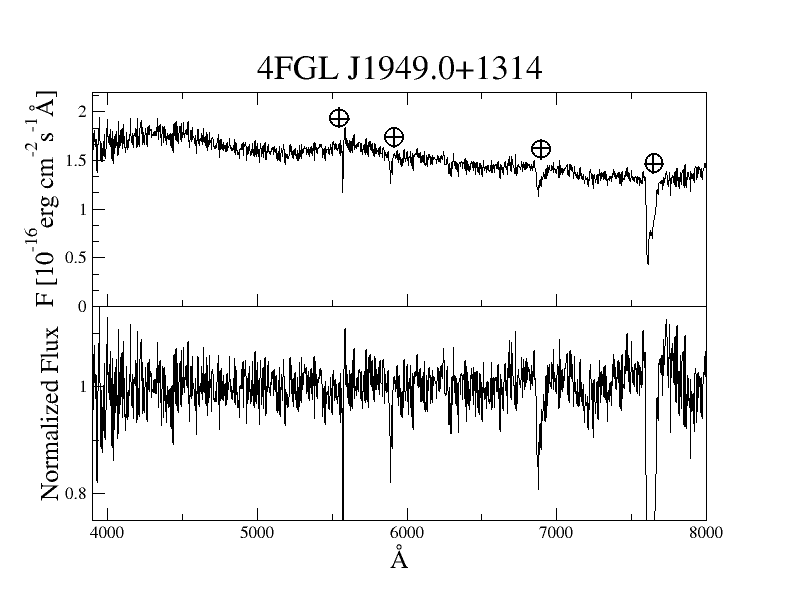}
	\caption{Same as Figure~\ref{fig:A1}.}
\end{figure*}


\bsp	
\label{lastpage}
\end{document}